# Higher borides and oxygen-enriched Mg-B-O inclusions as possible pinning centers in nanostructural magnesium diboride and the influence of additives on their formation.

Tatiana Prikhna,[a,*] Wolfgang Gawalek,[b] Yaroslav Savchuk,[a] Vasiliy Tkach,[a] Nikolay Danilenko,[c] Michael Wendt,[b] Jan Dellith,[b] Harold Weber,[d] Michael Eisterer,[d] Viktor Moshchil,[a] Nina Sergienko,[a] Artem Kozyrev,[a] Peter Nagorny,[a] Andrey Shapovalov,[a] Vladimir Melnikov,[a] Sergey Dub,[a] Doris Litzkendorf,[b] Tobias Habisreuther,[b] Christa Schmidt,[b] Athanasios Mamalis,[e] Vladimir Sokolovsky,[f] Vladimir Sverdun,[a] Fridrich Karau,[g] Alexandra Starostina[a]

[a]*Institute for Superhard Materials of the National Academy of Sciences of Ukraine, 2, Avtozavodskaya Str., Kiev 04074, Ukraine*

[b]*Institut für Photonische Technologien, Albert-Einstein-Strasse 9, Jena, D-07745, Germany*

[c]*Institute for Problems of Materialscience of the National Academy of Sciences of Ukraine, 3, Krzhyzhanivsky Str., Kiev 03142, Ukraine*

[d]*Atomic Institute of the Austrian Universities, 2, Stadionallee Str,. Vienna 1020, Austria*

[e]*National Technical University of Athens, 9, Iroon Polytechiou, Athens 15780, Greece*

[f]*Ben-Gurion University of the Negev, P.O.B. 653, Beer-Sheva 84105, Israel*

[g]*H.C. Starck GmbH, Goslar 38642, Germany*

**Abstract**

The study of high pressure (2 GPa) synthesized $MgB_2$-based materials allows us to conclude that higher borides (with near $MgB_{12}$ stoichiometry) and oxygen-enriched Mg-B-O inclusions can be pinning centers in nanostructural magnesium diboride matrix (with average grain sizes of 15–37 nm). It has been established that additions of Ti or SiC as well as manufacturing temperature can affect the size, amount and distribution of these inclusions in the material structure and thus, influence critical current density. The superconducting behavior of materials with near $MgB_{12}$ stoichiometry of matrix is discussed.
© 2001 Elsevier Science. All rights reserved

## 1. Introduction

The most apparent advantages of high–pressure synthesis of $MgB_2$ are the possibility to suppress the magnesium evaporation and formation of near theoretically dense nanostructural material with good connectivity between grains and high critical current density, $j_c$ in a short time (1 hour) [1]. It is considered that due to the comparatively large coherent length pining centers in $MgB_2$ can be grain boundaries, nanosized grains of secondary phases, and inhomogeneities of the structure. Because of this one can attain high $j_c$ in nanocrystalline material, the SC properties can be improved by alloying with nanosized homogeneously distributed additives. The atomic resolution study of the oxygen incorporation into bulk $MgB_2$ [2] shows that precipitates of Mg(B,O) of size 20–100 nm are formed by ordered occupation of boron lattice sites by oxygen atoms, while the basic bulk MgB$_2$ crystal structure and orientation are retained. The periodicity of the oxygen ordering is dictated by the oxygen concentration in the precipitates and primarily occurs in the (010) plane. The presence of these precipitates correlates well with an improved critical current density and superconducting transition behavior, implying that they act as pinning centers [2]. It has been established in our previous study that superconductive properties of MgB$_2$ depend on the amount, size, and distribution of higher borides inclusions with near MgB$_{12}$ stoichiometry (the finer the MgB$_{12}$ inclusions and the larger amount of them, the higher $j_c$) [3]. Besides, it has been shown that additions of Ti, Ta, and Zr can essentially improve $j_c$ of high-pressure high-temperature synthesized MgB$_2$, but we observed absolutely different mechanism of their influence than the one proposed for materials synthesized under pressureless conditions. In the case that Ti or Zr is added the improvement in critical current density in materials synthesized at ambient pressure is usually explained by the formation of TiB$_2$ or ZrB$_2$ thin layers or inclusions at grain boundaries that increase the number of pinning centers and hence the $j_c$ improvement caused by doping with these elements [4]. In the case of high-pressure synthesized MgB$_2$, the Ti-, Zr- or Ta-containing inclusions are rather coarse and randomly distributed in the material matrix to be pinning centers by themselves or to refine the MgB$_2$ structure. Under high-pressure conditions Zr, Ti or Ta absorb an impurity hydrogen (the source of which can be materials of high-pressure cell surrounding the sample during synthesis or admixture hydrogen in raw materials) to form TiH$_{1.94}$, ZrH$_2$, or Ta$_2$H, thus preventing harmful (for $j_c$) MgH$_2$ impurity phase from appearing and hydrogen from being introduced into the material structure. Besides, it has been observed that the presence, for example, of Ti or Ta promotes the formation of MgB$_{12}$ inclusions, which positively affect pinning in MgB$_2$-based materials, while the appearance of ZrB$_2$ in the structure does not affect the $j_c$ of high-pressure-synthesized (HPS) MgB$_2$-based ceramics. Here we discuss the structural inhomogeneities such as Mg-B-O inclusions and grains of higher borides in HPS MgB$_2$-based materials without and with Ti, Ta or SiC additions and their effect on SC characteristics. The revealed SC behavior of HPS materials having near MgB$_{12}$ stoichiometry of matrix phase is under consideration.

---

[*] Corresponding author. Tel.: +38-044-430-1126 , fax: +38-044-468-8625; e-mail: prikhna@mail.ru ; prikhna@iptelecom.net.ua .

## 2. Experimental

Samples were high-pressure (2 GPa) synthesized at 600 – 1050 °C from Mg and B in recessed-anvil high-pressure apparatuses [3] in contact with hexagonal BN. As the initial materials we used: powder of MgB$_2$ (H.C. Starck) with an average grain size of 10 $\mu$m and 0.8 % of O; several types of amorphous boron (H.C. Starck): B(I) 1.4 $\mu$m, 1.9% O, B(II) <5 $\mu$m 0.66 % O, B(III) 4 $\mu$m, 1.5 % O; metal magnesium chips (Technical Specifications of Ukraine 48-10-93-88), Ti (of size 1-3 $\mu$m, MaTecK, 99% purity), or SiC (200-800 nm, H.C. Starck). To produce MgB$_2$-based materials, metal magnesium chips and amorphous boron were taken in the stoichiometric ratio of MgB$_2$. To study the influence of Ti or SiC, (the?) powders were added to the stoichiometric MgB$_2$ mixture in amounts of 10 wt%. The components were mixed and milled in a high-speed activator with steel balls for 1-3 min and tabletized. To explore the processes of higher borides formation, Mg and B were taken in 1:4, 1:6, 1:7, 1:8, 1:10, 1:12, and 1:20 ratios and HPS at 800 and 1200 °C at 2 and 4 GPa for 1h.

The structure of materials was analyzed using TEM, SEM and X-ray diffraction. A scanning electron microscope ZEISS EVO 50XVP (resolution of 2 nm at 30 kV), equipped with: (1) an INCA 450 energy-dispersion analyzer of X-ray spectrums (OXFORD, England), using which the elements from boron to uranium can be quantitatively analyzed with a sensitivity of 0.1 wt %; a probe 2 nm in diameter; (2) a HKL Canell 5 detector of backscattering electrons (OXFORD, England), which allows us to get (using the Kikuchi method) the diffraction reflections of electrons from regions and layers of 10-1000 nm was employed. A JXA 88002 was used for SEM study. The microstructure analysis on the nanometer scale was carried out using JEM-2100F TEM equipped with an Oxford INCA energy detector with a probe of diameter 0.7 nm. Quantitative TEM-EDX analysis of boron was performed using the Oxford INCA energy program.

The $j_c$ was estimated by an Oxford Instruments 3001 vibrating sample magnetometer (VSM) using Bean's model. Hardness was measured employing a Matsuzawa Mod. MXT-70 microhardness tester, H$_V$ (using a Vickers indenter) and Nano-Indenter II, H$_B$ (using a Berkovich indenter).

## 3. Results and discussion

A material sintered from MgB$_2$ contained much less grains with near MgB$_{12}$ stoichiometry (Fig.1a, black grains) than that synthesized from Mg:B(I)=1:2 (Fig. 1b), which affects $j_c$ (Fig. 1d), but the presence of this phase is not

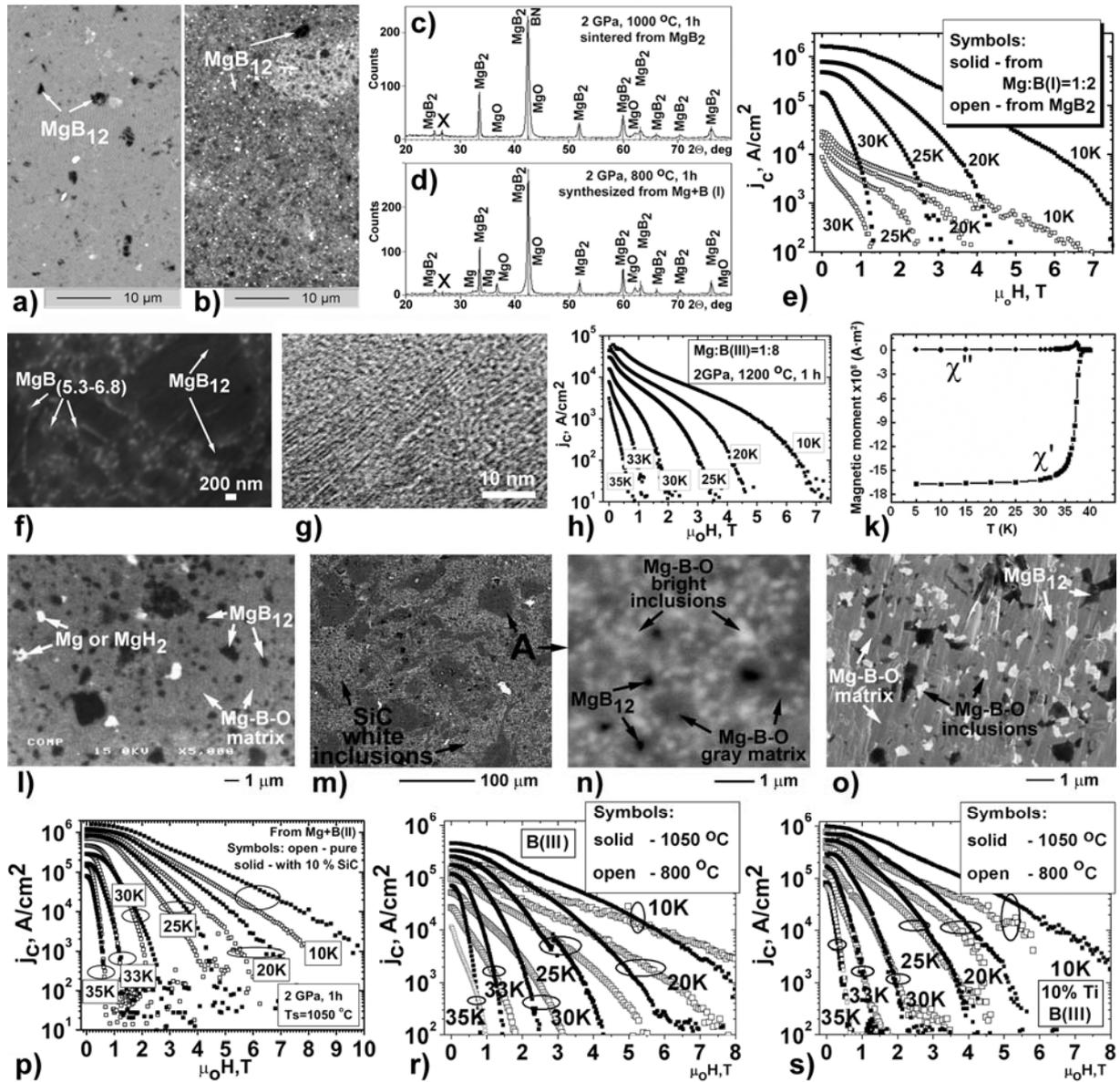

Fig.1 (a), (b) Images of the sample structure obtained by SEM in COMPOsitional contrast: (a) sintered from $MgB_2$ at 2 GPa, 1000 °C, 1 h; (b) synthesized from Mg and B taken into 1:2 ratio at 2 GPa, 800 °C; (c),(d) –X-ray patterns of the samples shown in Figs. 1a, b; (e) dependences of critical current density, $j_c$, on magnetic fields, $\mu_o H$ at different temperatures of the samples shown in Figs. 1a, b: open symbols indicate the sample sintered from $MgB_2$ material and solid symbols indicate the sample synthesized from Mg and B (I) taken into 1:2 ratio; (f–k) characteristics of the material synthesized from Mg and B(III) taken into 1:8 ratio at 2 GPa, 1200 °C, 1 h: (f), (g) high resolution SEM and TEM pictures, respectively, in COMPOsitional contrast (Fig.1 g demonstrates the structure of $MgB_{12}$ grain obtained by TEM), (h) dependences of $j_c$, on $\mu_o H$ at different temperatures, (k) imaginary ($\chi''$) and real ($\chi'$) part of the ac susceptibility (magnetic moment) vs. temperature, T, measured in ac magnetic field with 30 μT amplitude, which varied with a frequency of 33 Hz (k); (l-o) SEM pictures in COMPOsitional contrast of materials synthesized: (l), (o) from Mg:B(III) =1:2 at 2 GPa for 1 h at 800 °C without additions and at 1050 °C with 10% of Ti (the picture was taken at a place free from the Ti-containing inclusions ($TiH_2$)), respectively; (m),(n) from Mg:B(II)=1:2 with 10 % SiC addition under different magnifications; (p) dependences of $j_c$, on $\mu_o H$ at different temperatures of materials from Mg:B(II)=1:2 without additions (open symbols) and with 10 % of SiC added (solid symbols) HPSed at 2 GPa, 1050 °C for 1 h (structure of the sample synthesized at 1050 °C with 10 % of SiC is shown in Figs. 1m, n); (r), (s) dependences of critical current density, $j_c$, on magnetic fields, $\mu_o H$, at different temperatures for the materials synthesized from Mg and B(III) taken into 1:2 ratio at 2 GPa for 1 h (at 800 °C – open symbols and at 1050 °C – solid symbols; without additions (r) and with 10. % of Ti added (structure of the sample synthesized at 1050 °C with 10 % of Ti is shown in Fig. 1o).

reflected in X-ray pattern (Figs. 1c, d). The SEM study shows that the near-$MgB_2$-stoichiometry-matrix of HP-sintered sample (Fig. 1b) contains about 7 wt % O but that of the HP-synthesized one (Fig. 2a) in dark-gray places contains 10 wt% O and in light-gray places up to 25 wt% O. There is practically no oxygen (1-3 wt% O) in $MgB_{12}$ inclusions.

As estimated by X-ray analysis, the average grain size of material increases from 18–20 nm to 20–37 nm as sintering and synthesis temperature increase from 700 to 1000°C. Our multitudinous studies allow us to conclude that higher borides can be responsible for the reflex "x" in the X-ray pattern (this reflex corresponds exactly to hexagonal BN, which is used to protect samples from graphite heater, but the absence of nitrogen, according to the energy dispersive analysis, indicates that there is no BN in the material). To see higher borides in an X-ray pattern is impossible by several reasons: poor diffracted signals because of the low X-ray atomic scattering factor of boron [5], besides, $MgB_{12}$ inclusions (from 10μm to 20 nm) are dispersed in $MgB_2$, the etalon X-ray pattern of $MgB_{12}$ is absent in the database and the literature data are contradictory (the question which crystal symmetry (hexagonal or orthorhombic) has $MgB_2$ is still an open question), and the structure of $MgB_{12}$ grains (Fig. 1g) is far from perfection and contains many mosaic subgrains as the TEM study shows, and may be because of this we failed to get sharp reflections of Kikuchi lines from them.

The HP-synthesized samples from mixtures with great quantity of boron (up to Mg:B=1:20) exhibited SC properties [1], Fig. 1h, k and the highest $j_c$ as well as $T_c$ about 37 K were demonstrated by samples with near $MgB_{12}$ composition of matrix (prepared from the Mg:B= 1:8 and 1:20 mixtures). The estimated amount of shielding fraction in the sample (Figs. 1f–k) is 95.3% (this is a corrected value from the point of view of behavior of the sample with definite shape and sizes in magnetic field), which is indicate that a large volume of SC phase is present in the sample. As high-resolution TEM and SEM energy-dispersion analyses show (see Fig. 1f, g), the sample mainly contains phase with near the $MgB_{12}$ stoichiometry and some amount of $MgB_{(5.3-6.8)}$ phase is present as well. The $MgB_2$ phase was found in the form of randomly distributed inclusions of size less than 100 nm. The Berkovich nanohardness and Young modulus of the $MgB_{12}$ inclusions under 10–60 mn loads were 32.2±1.7 and 385±14 GPa, respectively, while for sapphire they were 31.1±2.0 and 416±22 GPa, respectively, and 17.4±1.1 and 213±18 GPa, respectively, for the matrix phase of the sample, in which $MgB_{12}$ inclusions were located. The Vickers microhardness under the 4.9 N-load of HPS sample with a matrix having the near $MgB_{12}$ stoichiometry (Fig. 1f) was 25.6±2.4 GPa, while of a sample with the near $MgB_2$ stoichiometry it 13.08±1.07 GPa. The SC properties of HPS samples with near $MgB_{16}$ composition of matrix were very poor.

Figs. 1p-s show the $j_c$ variations of $MgB_2$ obtained at different synthesis temperatures, $T_s$, from B and Mg taken in the 1:2 ratio with and without additions. With $T_s$ increasing (from 800 °C to 1050 °C, for example) for the materials with Ti (1o) or SiC (1m, n) additions the oxygen content of Mg-B-O inclusions increases, and the oxygen content of the materials matrix decreases (from 8 to 5 wt.%). While the incorporation of oxygen into Mg-B-O inclusions (with the Ts increase) is less pronounced in materials without additions. The formation of Mg-B-O inclusions and the decrease of oxygen content in material matrix are accompanied by an increase of $j_c$ at 10-25 K (Figs. 1p-s) in low and medium magnetic fields. Some decrease of $j_c$ observed in high magnetic fields in $MgB_2$-based materials obtained at higher $T_s$ correlates with the decrease of the $MgB_{12}$ volume in their matrix.

## 4. Conclusions

In the nanostructure of $MgB_2$-based materials high pressure synthesized *in-situ* from Mg and B the higher amount and more dispersed inclusions of $MgB_{12}$ phase are formed as compare to that high-pressure sintered *ex-situ* (from $MgB_2$). This is one of the factors, which allows higher critical current densities to be achieved. The Ti addition leads to an increase in the amount of $MgB_{12}$ in $MgB_2$ matrix. Besides, Ti and SiC contribute to the oxygen segregation and oxygen-enriched Mg-B-O inhomogeneities formation, which positively affects pining and $j_c$.

Materials with $MgB_{12}$ matrix with randomly distributed $MgB_2$ inclusions of about 100 nm exhibited SC behaviour.